\documentclass[aps,pra,twocolumn,longbibliography]{revtex4-1}
\usepackage{graphicx}
\usepackage{natbib}
\usepackage{amsmath}

\newcommand{\TJWat}{IBM T.J. Watson Research Center, Yorktown Heights, NY 10598, USA}

\newcommand{\figfolder}[1]{}

\usepackage[usenames,dvipsnames]{xcolor}
\usepackage[dvips,letterpaper=true,pagebackref=false]{hyperref}
\hypersetup{
	pdfnewwindow=true, 
	colorlinks=true, 
	linkcolor=MidnightBlue,
	citecolor=Magenta,
	urlcolor=RoyalBlue
}

\begin{document}

\title{Efficient Z-Gates for Quantum Computing}

\author{David C. McKay}
\email{dcmckay@us.ibm.com}
\author{Christopher J. Wood}
\author{Sarah Sheldon}
\author{Jerry M. Chow}
\author{Jay M. Gambetta}
\affiliation{\TJWat}
\date{\today}

\begin{abstract}
	For superconducting qubits, microwave pulses drive rotations around the Bloch sphere. The phase of these drives can be used to generate zero-duration arbitrary ``virtual'' Z-gates which, combined with two $X_{\pi/2}$ gates, can generate any SU(2) gate. Here we show how to best utilize these virtual Z-gates to both improve algorithms and correct pulse errors. We perform randomized benchmarking using a Clifford set of Hadamard and Z-gates and show that the error per Clifford is reduced versus a set consisting of standard finite-duration X and Y gates. Z-gates can correct unitary rotation errors for weakly anharmonic qubits as an alternative to pulse shaping techniques such as DRAG. We investigate leakage and show that a combination of DRAG pulse shaping to minimize leakage and Z-gates to correct rotation errors (DRAGZ) realizes a 13.3~ns $X_{\pi/2}$ gate characterized by low error ($1.95[3]\times 10^{-4}$) and low leakage ($3.1[6]\times 10^{-6}$). Ultimately leakage is limited by the finite temperature of the qubit, but this limit is two orders-of-magnitude smaller than pulse errors due to decoherence. 
\end{abstract}

\maketitle

Computers based on quantum bits (qubits) are predicted to outperform classical computers for certain critical problems, e.g., factoring. Unlike a classical bit, which is discretely in the state 0 or 1, a qubit can be in a superposition state $|\Psi\rangle=\cos(\theta/2)|0\rangle + e^{i\phi}\sin(\theta/2)|1\rangle$  where $|0\rangle$ and $|1\rangle$ are the quantum versions of the classical 0 and 1 states. This single-qubit superposition state can be geometrically represented as a point on the surface of a  unit-sphere known as the \emph{Bloch sphere}. Critical to implementing a quantum computer is the ability to control the state of the qubit, i.e., transform the qubit state arbitrarily between two points on the Bloch sphere. This is accomplished by unitary transformations (gates), which correspond to rotations of the state around different axes in the Bloch sphere representation. Physically, X and Y gates (rotations around the X and Y axes) are generated by modulating the coupling between the states $|0\rangle$ and $|1\rangle$ at the frequency difference between these states $\omega_{01}=(E_{|1\rangle}-E_{|0\rangle})/\hbar$. This modulation drive has the general form $\Omega(t) \cos(\omega_{D} t - \gamma)$ where $\Omega(t)$ is the drive strength of the rotation, $\omega_D$ is the drive frequency ($\omega_D=\omega_{01}$ on resonance) and $\gamma$ is the drive phase. The duration of the gate is set by the desired rotation angle and the drive strength. On-resonance, when $\gamma=0$, the qubit state rotates around the X axis and when $\gamma=\frac{\pi}{2}$ the rotation is around the Y axis. Therefore, the geometric X and Y axes in the Bloch sphere correspond to a real $\frac{\pi}{2}$ phase difference between drive fields.

 Rotations around the remaining axis (Z axis), i.e., Z-gates, correspond to a change in the relative phase between the $|0\rangle$ and $|1\rangle$ states. A Z-gate can be implemented by either detuning the frequency of the qubit with respect to the drive field for some finite amount of time (e.g. see Ref.~\cite{lucero:2010}) or by composite X and Y gates. The result is that the qubit state rotates with respect to the X and Y axes. However, it is equivalent to rotate the axes with respect to the qubit state -- such a gate is known as a virtual Z-gate which corresponds to adding a phase offset to the drive field for all subsequent X and Y gates. This includes adding a phase offset to any two-qubit drives such as a drive used to implement a CNOT gate via cross-resonance~\cite{chow:2011}. In many qubit implementations this phase, $\gamma$, is defined by classical control hardware and software. A Z-gate implemented in this way is essentially perfect; the classical hardware is self-calibrated via a global frequency reference (e.g., an atomic clock) and the gate has zero duration. Virtual-Z (VZ) gates have long been used in quantum experiments such as in NMR~\cite{knill:2000}, ions~\cite{knill:2008}, and superconducting qubits~\cite{johnson:2015}. Utilizing these gates can improve the overall fidelity of a quantum circuit if the circuit is optimized to maximize the number of single-qubit Z-gates. Additionally, any arbitrary rotation in the Bloch sphere can be generated by combining Z and $X_{\pi/2}$ gates. This greatly simplifies calibration procedures because only a single drive strength must be calibrated. 

Furthermore, Z-gates can compensate for certain unitary errors which occur in physical qubit implementations. For example, when driving X and Y rotations in weakly anharmonic superconducting transmon qubits~\cite{koch:2007} there are unitary rotation errors (Stark shifts errors) and population leakage. Both of these errors can be corrected by implementing a full DRAG pulse~\cite{motzoi:2009}, which involves pulse shaping and dynamic frequency tuning. However, only the pulse shaping component of DRAG is typically implemented~\cite{chow:2010,lucero:2010,gambetta:2011}, which cannot simultaneously correct both errors (we herein refer to DRAG by this definition). In most experiments DRAG is optimized to correct the more dominant unitary errors. This problem is solved by adding VZ-gates since VZ-gates can correct unitary phase errors while pulse shaping is then optimized to minimize leakage. Similar errors are common when driving two-qubit gates, such as the parametric iSWAP~\cite{mckay:2016}, cross-resonance~\cite{sheldon:2016} and adiabatic CZ~\cite{dicarlo:2009}, so VZ-gates plus pulse shaping is also applicable in multiqubit systems.

In this paper we explore how the VZ-gate can be used to minimize circuit error and, for superconducting transmon qubits, minimize pulse errors. First, we review the theory of the VZ-gate and show one specific formula for an arbitrary SU(2) gate. Next, we compare randomized benchmarking~\cite{magesan:2011} of a qubit using Cliffords generated from two different sets of basis gates -- one set using X and Y rotations and the other set using X and Z rotations.  We show that the error per Clifford is lower for the XZ basis set, since the the number of finite-duration gates (i.e., X and Y rotations) required to represent each Clifford gate is reduced. We also perform interleaved benchmarking of the $S$ gate ($Z_{\pi/2}$) and measure an error rate that is consistent with a perfect gate. Next, we demonstrate that VZ-gates can be used to compensate Stark-shift errors that arise when driving most X and Y rotations in weakly anharmonic systems. This technique, Gaussian plus Z (GZ), is a straightfoward alternative to the commonly used DRAG pulse~\cite{motzoi:2009,chow:2010,lucero:2010,gambetta:2011}. We show via randomized benchmarking that the pulse error for DRAG and GZ is essentially equivalent. Next, we measure population leakage to the $|2\rangle$ state during these RB sequences (see Ref.~\cite{chen:2016} for similar work). We show that GZ and DRAG (optimized for fidelity) have similar leakage rates. Combining DRAG and VZ-gates (DRAGZ) improves leakage without a loss of fidelity as does passive filtering plus VZ-gates (FILTZ). For pulses longer than 25~ns we find that the leakage rates for all methods are similar and limited by heating. Finally, we discuss some considerations when using these gates in a multi-qubit system. 

\section{Theory}

\begin{figure}
\includegraphics[width=0.4\textwidth]{\figfolder{1a}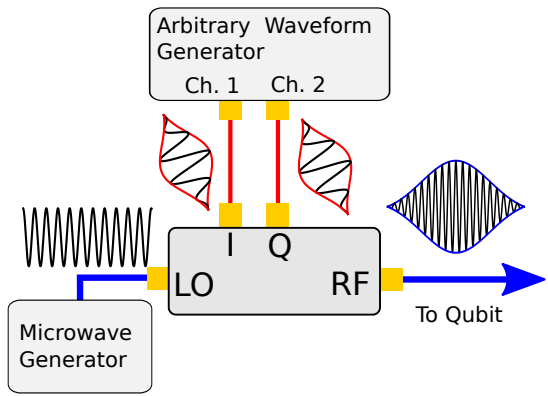}
\caption{(Color Online) Schematic of the typical experimental setup for generating shaped microwave pulses for driving superconducting qubits. \label{fig:1a}}
\end{figure}

To elucidate the concept of the virtual Z-gate (VZ-gate), we first review basic single qubit gates as they are physically realized in many labs, i.e., with a local oscillator (LO) shaped by an arbitrary waveform generator (AWG) through an IQ mixer (setup shown in Fig.~\ref{fig:1a}). The LO is a single tone microwave source that outputs a constant signal $\cos(\omega_{LO}t)$. The AWG outputs a programmable series of discrete voltage points at a specific sample rate, e.g., for the Tektronix 5014 used in these experiments that sample rate is 1.2GSa/s (0.833ns between points). These points are internally filtered so that the output is a smooth waveform. Including this filter ($F$), the AWG output is
\begin{equation}
	V(t) = \int_{0}^{t} d\tau F(t-\tau) \sum_{n=0}^{\infty} V_n \sqcap(n-\tau/T) \label{eqn:awgfilt}
\end{equation}
where $\{V_n\}$ is the set of AWG voltages, $T$ is the AWG period and $\sqcap(x)$ is a pulse function defined as $\sqcap(x)=\{1,0\}$ conditioned on $|x|\le 1/2$. The IQ mixer multiplies the I and Q channels with the LO such that,
\begin{eqnarray}
	V_{RF}(t) & = & V_{I}(t)\cos(\omega_{LO}t)  \nonumber \\
			 & & + V_{Q}(t)[1+\epsilon_Q]\sin(\omega_{LO}t+\epsilon_{\phi}) \nonumber \\
			 & & + \epsilon_{LO}\cos(\omega_{LO} t)
\end{eqnarray}
where the $\epsilon$ terms are non-ideal errors common to IQ mixers. 

The AWG voltages are used to shape the pulse and shift the pulse frequency to resonance via single sideband (SSB) modulation. If the desired pulse envelope is $\Omega(t)$ and the drive frequency is $\omega_{D}=\omega_{LO}+\omega_{SSB}$ then the output of an ideal AWG (ignoring the pixelation and filtering described by Eqn.~\ref{eqn:awgfilt}) is, 
\begin{eqnarray}
	V_{I}(t) & = & \Omega(t) \cos(\omega_{SSB} t - \gamma), \\
	V_{Q}(t) & = & -\Omega(t) \sin(\omega_{SSB} t - \gamma).
\end{eqnarray}
These AWG signals, when applied to the inputs of an ideal mixer, output the desired drive pulse
\begin{eqnarray}
	V_{RF}(t) & = & \Omega(t) \cos(\omega_{D} t - \gamma). \label{eqn:pulse_drive}
\end{eqnarray}
A series of $n$ shaped microwave pulses driving an anharmonic oscillator (a good description of a transmon qubit) is described by the Hamiltonian (in the lab frame),
\begin{eqnarray}
	H/\hbar & = & \sum_{n} \Omega_{n}(t) \cos(\omega_D t +\gamma_n) (\hat{a}+\hat{a}^{\dagger}) + \omega_{01} \hat{n} \nonumber \\
	& & + \frac{\alpha}{2} (\hat{n}-1)\hat{n},
\end{eqnarray}
where $\hat{a}(\hat{a}^{\dagger})$ is the annihilation (creation) operator of the oscillator, $\hat{n}=\hat{a}^{\dagger}\hat{a}$ is the number operator and $\alpha$ is the anharmonicity. The $|0\rangle$ and $|1\rangle$ levels of this oscillator are the qubit levels. Leakage to higher levels and unitary errors due to mixing with these levels will be discussed in \S~\ref{sect:err} and \ref{sect:leakage}. For simplicity we rewrite just the qubit Hamiltonian,
\begin{equation}
	H/\hbar = \sum_{n} \Omega_{n}(t) \cos(\omega_D t - \gamma_n) \hat{\sigma}_X - \frac{\omega_{01}}{2} \hat{\sigma}_Z,
\end{equation}
where $\hat{\sigma}_{X},\hat{\sigma}_{Z}$ are the Pauli operators. For a resonant drive ($\omega_D=\omega_{01}$), the Hamiltonian in the qubit rotating frame is,
\begin{equation}
	\tilde{H}/\hbar = \sum_{n} \frac{\Omega_{n}(t)}{2}  \left[ \cos(\gamma_n) \hat{\sigma}_X + \sin(\gamma_{n}) \hat{\sigma}_Y\right] \end{equation}
Assuming a constant amplitude pulse $\Omega_{n}$ for duration $T$, the unitary transformation (the gate) due to pulse $n$ is,
\begin{equation}
	U_n = e^{-i\frac{\Omega_n T}{2} \left[ \cos(\gamma_n) \hat{\sigma}_X + \sin(\gamma_n) \hat{\sigma}_Y\right]}, \label{eqn:u_rot}
\end{equation}
and so a $\gamma_n=0 (\frac{\pi}{2})$ pulse is a rotation of angle $\Omega_n T$ around the X (Y) axis of the Bloch sphere. Therefore, the AWG controls both the pulse rotation and the rotation axes. Since $\gamma$ controls the rotation axes, intuitively we can perform a VZ-gate by adjusting $\gamma$. More formally, we can show how this works by consider two consecutive $X$ pulses, $X_{\theta}=e^{-i\frac{\theta}{2} \hat{\sigma}_X}$, with the phase $\gamma$ offset by $\phi$ between the pulses. The total unitary is,
\begin{equation}
	e^{-i\frac{\theta}{2} (\cos(\phi)\hat{\sigma}_X + \sin(\phi) \hat{\sigma}_Y)} \cdot X_{\theta},
\end{equation}
which can be expanded to,
\begin{equation}
e^{i\frac{\phi}{2} \hat{\sigma}_Z} \cdot e^{-i\frac{\theta}{2}\hat{\sigma}_X} \cdot e^{-i\frac{\phi}{2}\hat{\sigma}_Z} \cdot X_{\theta},
\end{equation}
which equals
\begin{equation}
	Z_{-\phi} \cdot X_{\theta} \cdot Z_{\phi} \cdot X_{\theta}.
\end{equation}
Therefore, by simply adding a phase offset in software and redefining the rotation axes for subsequent X and Y gates, we can effectively implement an arbitrary Z-gate. The additional $Z_{-\phi}$ gate at the end is due to the fact that we are in the qubit frame of reference and so the phase offset $\phi$ must be carried through for all subsequent gates. For example, if we follow our original sequence with a $Y_{\theta}$ gate with the phase offset applied the gate sequence is
\begin{eqnarray}
	& &  e^{-i\frac{\theta}{2} \left(\cos\left[\frac{\pi}{2}+\phi\right]\hat{\sigma}_X + \sin\left[\frac{\pi}{2}+\phi\right] \hat{\sigma}_Y\right)} \cdot \ldots \nonumber \\
	  & & Z_{-\phi} \cdot X_{\theta} \cdot Z_{\phi} \cdot X_{\theta}, \\
	  & = &  Z_{-\phi} \cdot Y_{\theta} \cdot Z_{\phi} \cdot   Z_{-\phi} \cdot X_{\theta} \cdot Z_{\phi} \cdot X_{\theta}, \\
	& = &  Z_{-\phi} \cdot Y_{\theta} \cdot  X_{\theta} \cdot Z_{\phi} \cdot X_{\theta}.
\end{eqnarray}
The inverse Z-gate remains, but does not change the measurement outcomes which are measured along Z. 

Given that we can easily create Z-gates, we now show that any arbitrary SU(2) gate can be constructed by combining Z-gates with two $X_{\pi/2}$ gates. In general, any SU(2) gate can be written in the form,
\begin{equation}
U(\theta,\phi,\lambda) = \left[\begin{array}{cc} \cos(\theta/2) & -i e^{i\lambda}\sin(\theta/2) \\ -ie^{i \phi} \sin(\theta/2) & e^{i(\lambda+\phi)} \cos(\theta/2) \end{array}\right]
\end{equation}
which is conveniently represented (up to a global phase) as,
\begin{equation}
U(\theta,\phi,\lambda) = Z_{\phi} \cdot X_{\theta} \cdot Z_{\lambda}.
\end{equation}
By using the identity,
\begin{equation}
	X_{\theta} = Z_{-\pi/2} \cdot X_{\pi/2} \cdot Z_{\pi-\theta} \cdot X_{\pi/2} \cdot Z_{-\pi/2},
\end{equation}
we show that any SU(2) gate is
\begin{equation}
U(\theta,\phi,\lambda) = Z_{\phi-\pi/2} \cdot X_{\pi/2} \cdot Z_{\pi-\theta} \cdot X_{\pi/2} \cdot Z_{\lambda-\pi/2}.
\end{equation}
We express some common gates in this notation in Table~\ref{tab:gates}. The ability to efficiently create arbitrary SU(2) gates using Z-gates is essential for performing universal quantum algorithms. 

\begin{table}
	\caption{\label{tab:gates}Common SU(2) gates expressed with Z gates.}
\begin{ruledtabular}
\begin{tabular}{llll}
	Gate & $\theta$ & $\phi$ & $\lambda$ \\ \hline
	$\mathcal{I}$ & 0 & 0 & 0  \\
	$X_{\pi}$ & $\pi$ & 0 & 0 \\ 
	$Y_{\pi}$ & $\pi$ & $\pi/2$ & -$\pi/2$ \\ 
	$Z_{\pi}$ & 0 & $\pi/2$ & $\pi/2$ \\ 
	$X_{\pi/2}$ & $\pi/2$ & 0 & 0 \\
	$Y_{\pi/2}$ & $\pi/2$ & $\pi/2$ & -$\pi/2$ \\
	$S$ & 0 & $\pi/4$ & $\pi/4$ \\
	$H$ & $\pi/2$ & $\pi/2$ & $\pi/2$ \\
	$X_{\pi/4}$ & $\pi/4$ & 0 & 0 \\
	$T$ & 0 & $\pi/8$ & $\pi/8$
\end{tabular}
\end{ruledtabular}
\end{table}

\section{Randomized Benchmarking of the Z-Gate \label{sect:zgates}}

To demonstrate how the VZ-gate can improve algorithms we perform randomized benchmarking~\cite{magesan:2011} (RB) using a fixed-frequency superconducting transmon qubit of frequency $\omega/2\pi=5.0353$~GHz, anharmonicity $\alpha/2\pi = -235.5$~MHz, and typical coherences $T_1=54(1)~\mu\mathrm{s}$, $T_{\phi}=135(4)~\mu\mathrm{s}$. This qubit is part of a two-qubit device detailed in Ref.~\cite{mckay:2016}; for the work here we consider it as a single independent qubit (the other qubit frequency is 5.924~GHz). A RB circuit consists of $m$ random Clifford gates with a final inverting gate so that the full circuit implements the identity operator. These Clifford gates are constructed from single qubit gate primitives which, at minimum, are $\frac{\pi}{2}$ pulses along two independent axes. Here we compare two sets of basis pulses --- the $XY_{\frac{\pi}{2}}$ and $HZ$ sets. The $XY_{\frac{\pi}{2}}$ set consists of the finite-duration gates $\{X_{\pi/2},X_{-\pi/2},Y_{\pi/2},Y_{-\pi/2}\}$ whereas the $HZ$ set consists of one finite duration gate combined with VZ-gates $\{H,I=Z_{0},S=Z_{\frac{\pi}{2}},S^{\dagger}=Z_{-\frac{\pi}{2}},Z_{\pi}\}$ where $H=Z_{\frac{\pi}{2}}\cdot X_{\frac{\pi}{2}}\cdot Z_{\frac{\pi}{2}}$ is the Hadamard gate. On average, 2.25 gates from the $XY_{\frac{\pi}{2}}$ set and 2.4583 gates from the $HZ$ set are required to construct a Clifford. However, for the HZ set only one of those gates per Clifford is the finite-duration Hadamard gate. Since we expect the VZ-gates to be near perfect, the HZ set should have lower error per Clifford. For this experiment each of these finite duration gates is implemented as a DRAG pulse --- a pulse with a Gaussian envelope along the main rotation axis and a pulse with a derivative Gaussian envelope along the orthogonal rotation axis. The Gaussian pulse is defined as,
\begin{eqnarray}
\Omega_{G}(t) & = & \left\{ \begin{array}{ccc} \Omega_0 \frac{e^{-t^2/2\sigma^2}-e^{-T^2/2\sigma^2}}{1-e^{-T^2/2\sigma^2}} &, & t \le T \\ 0 & , & \mathrm{else} \end{array} \right. \label{eqn:gauss}
\end{eqnarray}
where $T$ is the pulse length which is set to $T=4\sigma$. For these experiments $T=13.33$~ns and $\sigma=3.33$~ns with a $6.7$~ns buffer between pulses ($\omega_{SSB}/2\pi=-120$~MHz). The total DRAG pulse is then
\begin{eqnarray}
\Omega(t) & = & \left\{ \begin{array}{ccc} \Omega_{G}(t) & , & \gamma=0 \\ \beta \dot{\Omega}_{G}(t) & , & \gamma=\frac{\pi}{2} \end{array} \right., \label{eqn:drag}
\end{eqnarray}
where the DRAG parameter $\beta$ is calibrated to optimize pulse fidelity by cancelling Stark shift errors due to off-resonance driving of higher transmon levels. In theory, the value of $\beta$ which optimizes fidelity is $1/2\alpha$~\cite{gambetta:2011}, however, in practice the experimentally optimized value of $\beta$ is different since the DRAG pulse also compensates phase errors from other sources. DRAG pulses can also minimize population leakage to these higher levels, but the value of $\beta$ for which the pulse minimizes leakage is not generically the value for which the pulse maximizes fidelity. This point will be discussed in more detail in \S~\ref{sect:leakage} and is also addressed in Ref.~\cite{chen:2016}.

\begin{figure}
\includegraphics[width=0.45\textwidth]{\figfolder{1}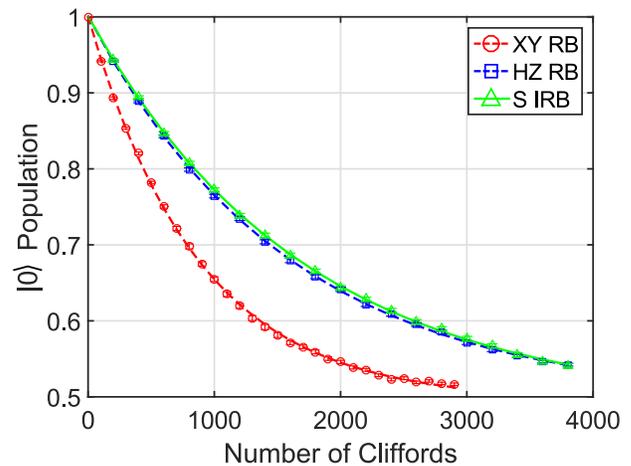}
\caption{(Color Online) RB curves for the two basis sets discussed in the main text: $XY_{\frac{\pi}{2}}$  (red circles) and $HZ$ (blue squares). Interleaved RB for the $S$ gate using the $HZ$ set (green triangles). Each point is the average of 20 random seeds run in 5 separate experiments  and fit to the standard RB exponential decay curve $A r^m + B$ where $m$ is the number of Clifford gates and the average error per Clifford is $\frac{1}{2}(1-r)$. The average error per gate is $\frac{1}{2}(1-r^{1/N_{g}})$ where $N_g$ is the number of gates in the basis set required to implement a Clifford~\cite{magesan:2011}. \label{fig:1}}
\end{figure}

Sample RB data is shown in Fig.~\ref{fig:1}. For the  $XY_{\frac{\pi}{2}}$ set, averaging over 5 runs of 20 seeds each, we get an error per Clifford (EPC) of 5.6(1)$\times10^{-4}$ and an error per gate in the set (EPG) of 2.48(5)$\times10^{-4}$. For the $HZ$ set, averaging over 10 runs, the EPC is 3.0(1)$\times10^{-4}$ and the EPG is 1.22(4)$\times10^{-4}$. The lower EPC of the $HZ$ set is evident from the data in Fig.~\ref{fig:1}, and confirms our expectation that the VZ-gates are significantly better than finite-length X and Y gates. To quantify the error of the VZ-gates we perform interleaved RB~\cite{magesan:2012} of the $S$ gate; sample data is shown in Fig.~\ref{fig:1}. Averaging over 5 runs we get an error of -1.7(1.0)$\times10^{-5}$ with systematic errors bounds of [0,6$\times10^{-4}$]. This is consistent with the VZ-gate having zero error and, therefore, being a perfect gate. 

The RB data demonstrate the advantage of utilizing the VZ-gates in quantum algorithms. In essence, each of the curves is implementing the same RB algorithm, i.e., generate a sequence of $m$ random Cliffords that constructs the identity operator. By utilizing VZ-gates we are able to implement the algorithm with higher fidelity. Therefore, VZ-gates can lower error rates in many algorithms by optimizing the circuit to mazimize the number of Z-gates. \\

\section{Correcting Errors with VZ-Gates \label{sect:err}}

\begin{figure}
\includegraphics[width=0.45\textwidth]{\figfolder{2a}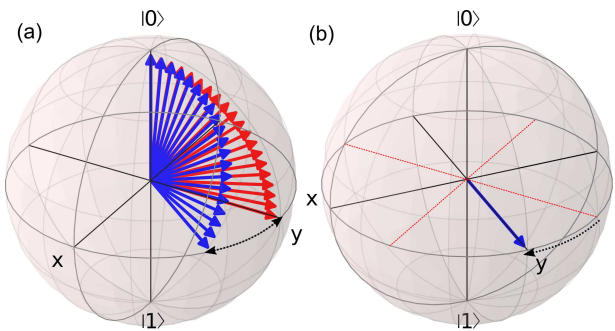}
\caption{(Color Online) (a) Bloch-sphere representation of an attempted $\frac{\pi}{2}$ rotation around the $X$ axis starting in the state $|0\rangle$ on-resonance (red/light gray) and with a detuned drive (blue/dark gray). For a suitably compensated rotation angle the detuned drive ends up in the $XY$ place, but with a finite Z-rotation error. (b) Correcting the error using a VZ-gate, i.e., axes rotation. \label{fig:2a}}
\end{figure}

Beyond their direct use in quantum circuits, VZ-gates can correct for certain pulse errors that occur during physical one and two-qubit gates without introducing new errors. For example, a VZ-gate can correct a phase error, i.e. an unwanted Z-gate, by applying the inverse Z-gate. VZ-gates can also correct most off-resonance-rotation (ORR) errors. The unitary operator due to an ORR along the $X$ axis is,
\begin{eqnarray}
	U_1 & = & e^{-it  \left[\frac{\Omega}{2}\hat{\sigma}_X + \Delta \hat{\sigma}_Z \right]}, \\
	U_1 & = & e^{-i\frac{\Omega_R t}{2} \left[\cos(\lambda)\hat{\sigma}_X + \sin(\lambda)\hat{\sigma}_Z \right]}, \label{eqn:rot_err_u1}
\end{eqnarray}
where $\tan(\lambda)=\frac{\Delta}{\Omega}$ and $\Omega_R=\sqrt{\Omega^2+\Delta^2}$. If the goal is to implement the gate $X_{\theta}$, then the question is whether there is a Z correction that can be applied to Eqn.~\ref{eqn:rot_err_u1}, i.e., is there a $\xi$ and $\Omega_R$ such that
\begin{equation}
	X_{\theta} = Z_{\xi} \cdot U_1(\Omega_R t,\lambda) \cdot Z_{\xi}? \label{eqn:rot_err_u2}
\end{equation}
If we expand Eqn.~(\ref{eqn:rot_err_u2}) then we get the following relations,
\begin{eqnarray}
	\sin\left(\frac{\Omega_R t}{2}\right) & = & \frac{\sin\left(\frac{\theta}{2}\right)}{\cos(\lambda)}, \label{eqn:omega}\\
	\tan(\xi) & = & \sin(\lambda) \tan\left(\frac{\Omega_R t}{2}\right), \label{eqn:xi}
\end{eqnarray}
and so there is a valid correction for the ORR error when $\frac{\sin\left(\frac{\theta}{2}\right)}{\cos(\lambda)}\le1$. For example, if  $\lambda=0.1$, and the desired gate is $X_{\frac{\pi}{2}}$ then $\xi \approx 0.1$ and $\Omega_R t= \pi/2+0.1^2$. Graphically, an exagerrated ORR for $X_{\pi/2}$ is illustrated on the Bloch sphere in Fig.~\ref{fig:2a}. Starting from $|0\rangle$ the rotation is off-axis and so the final state is not along the $Y$ axis. However, a rotation angle exists so that the state still crosses the $XY$ plane and then a final VZ-gate corrects the angle error. Physically there is no solution when the rotation is sufficiently off-resonance such that it cannot pass through the plane defined by the desired final state. For example, a detuned $\pi$ pulse cannot be compensated since a qubit starting in state $|0\rangle$ does not complete the rotation to $|1\rangle$.

\begin{figure}
\includegraphics[width=0.4\textwidth]{\figfolder{2}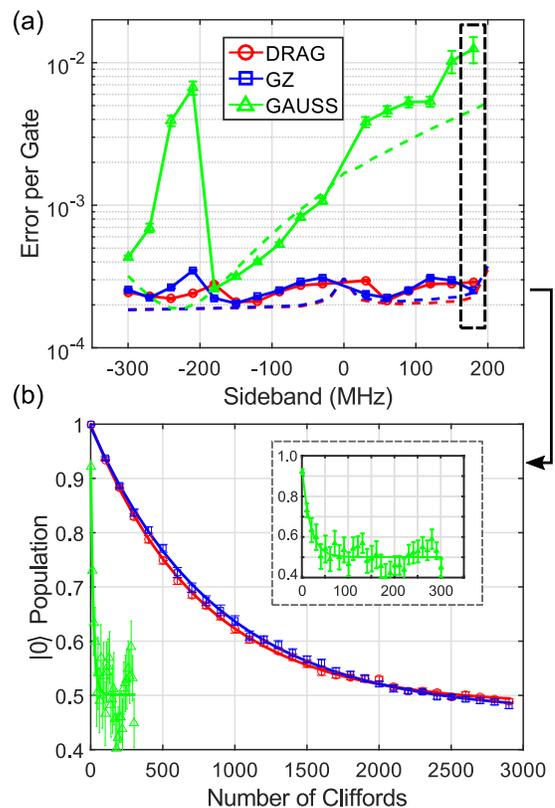}
\caption{(Color Online) (a) EPG for the $XY_{\frac{\pi}{2}}$ set (see main text) for three different pulse implementations: Gaussian, DRAG and Gaussian plus VZ-gate (GZ) as a function of the sideband frequency. Dashed lines are theory fits assuming the coherence numbers listed in the main text, LO leakage of -65~dBm and an internal AWG filter approximately as a Gaussian filter with a bandwidth of 300~MHz\cite{motzoi:2011}. (b) Sample RB curves for the different pulses for a sideband frequency of 180~MHZ. The DRAG and GZ pulses are coherence limited, but the Gaussian pulse is completely dominated by unitary errors.  \label{fig:2}}
\end{figure}

Correcting ORR errors is of practical importance for weakly anharmonic transmon qubits. When resonantly driving the $|0\rangle$ to $|1\rangle$ transition, the drive frequency is only slightly detuned from the higher level transitions such as $|1\rangle$ to $|2\rangle$, and so there is a strong Stark effect which shifts the frequency of the $|1\rangle$ state during the drive. The strength of the Stark shift is inversely proportional to the detuning, and thus ORR errors increase for short gates because of Fourier broadening of the drive frequency. The standard approach to correct these errors is to utilize DRAG pulse shaping, Eqn.~(\ref{eqn:drag}), as we did for the RB data in \S~\ref{sect:zgates}. Here we show similar performance using Gaussian pulses with VZ-gates used to correct ORR errors in the form given by Eqn.~(\ref{eqn:rot_err_u2}); we refer to the combined pulse as Gaussian plus Z (GZ). In Fig.~\ref{fig:2} we plot the error per gate (EPG) from $XY_{\frac{\pi}{2}}$ RB for DRAG, GZ and Gaussian pulses versus the sideband frequency of the pulse. Interestingly, the EPG for the Gaussian pulse is a strong function of the sideband frequency. This can be understood to be the result of the internal filtering of the AWG. In the rotating frame of the qubit the effect of this filter on the pulse shape is similar to the DRAG pulse shape. The value and sign of $\beta$ is a function of the sideband frequency and so for certain frequencies the pulse shape exacerbates the ORR error. When we apply the Z gate correction the calibrated phase compensates for any passive DRAG shaping. From this data we conclude that GZ pulses are a viable alternative to DRAG when optimizing pulse fidelity. GZ pulses are not sensitive to the exact shape of the pulse and, as we will discuss next, permit the utilization of pulse shaping to address different errors such as leakage.  

\section{Leakage \label{sect:leakage}}

In addition to unitary ORR errors, there is also population leakage to higher levels when driving transmon qubits. This leakage is mainly caused by frequency components in the drive at the transition frequency between the $|1\rangle$ and $|2\rangle$ states, $\omega_{12}=\omega_{01}+\alpha$. Pulse shaping, i.e., DRAG, can effectively mitigate leakage when VZ gates can be utilized to correct the unitary ORR errors (see Ref~\cite{chen:2016} for similar work using DRAG and frequency chirped pulses). Here we analyze leakage versus pulse width for different pulse types: DRAG optimized for fidelity, GZ, DRAG optimized for leakage with VZ-gates to correct ORR errors (DRAGZ) and GZ with the AWG outputs externally low-pass filtered (FILTZ). For these leakage experiments we operate with a sideband frequency of -120MHz, which has two advantages. For one, leakage components at $\omega_{\mathrm{LO}}$ and $\omega_{\mathrm{LO}}-\omega_{\mathrm{SSB}}$ due to non-idealities in the mixer are detuned by at least $|\alpha+\omega_{\mathrm{SSB}}|$ from $\omega_{12}$. Second, by selecting a negative sideband we can passively filter the AWG for signals at $|\alpha+\omega_{\mathrm{SSB}}|$ which could mix with $\omega_{\mathrm{LO}}$ to produce $\omega_{12}$. For the qubit in these experiments $|\alpha+\omega_{\mathrm{SSB}}|/2\pi= 355.5$~MHz, so a LP filter between 120~MHz and 355.5~MHz can effectively filter leakage components and leave enough bandwidth to drive short pulses. Specifically, we filtering using Mini-Circuits VLF-180 (3dB frequency of 270~MHz). By filtering the AWG output we are effectively implementing a passive form of pulse shaping. 

\begin{figure}
\includegraphics[width=0.35\textwidth]{\figfolder{3a}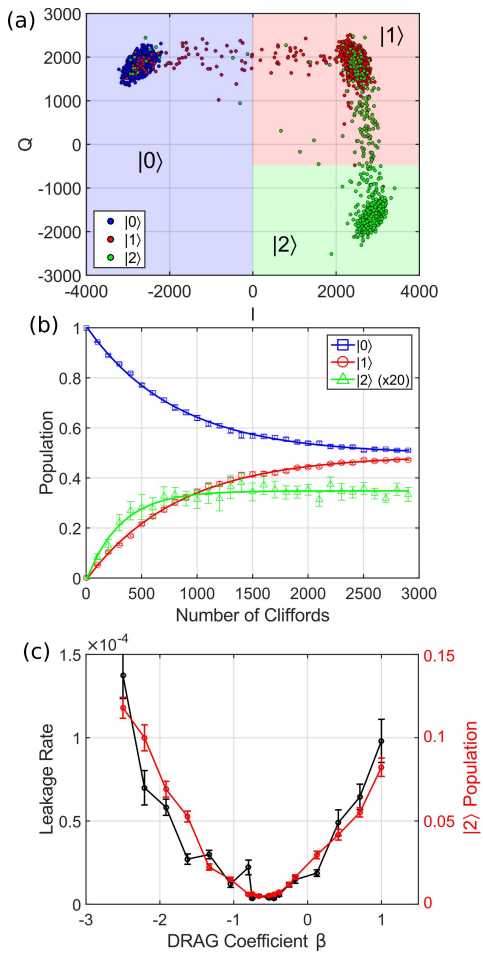}
\caption{(Color Online) (a)Single-shot measurements of the $|0\rangle$ (blue, left),$|1\rangle$ (red, top-right) and $|2\rangle$ (green, bottom-right) states in the IQ plane (1000 shots). The colored regions demonstrate the hard thresholding barriers used to bin measurements. For the RB data we use the results from this calibration to construct a POVM and invert the hard-threshold results to compensate for assignment errors. (b) Typical RB measurement for leakage. The $|2\rangle$ state population is multiplied by 20 times for scale.  (c) Leakage rate and $|2\rangle$ population after 2901 Clifford gates (averaged over 20 seeds) versus the DRAG parameter $\beta$ (black). This $|2\rangle$ population is used to calibrate $\beta$ for the DRAGZ pulses (red/light gray).    \label{fig:3a}}
\end{figure}

To measure leakage we perform standard RB sequences and measure the $|0\rangle$, $|1\rangle$ and $|2\rangle$ populations simultaneously by hard thresholding single-shot readout signals as shown in Fig.~\ref{fig:3a} (a). A sample RB curve for the 3 states is shown in Fig.~\ref{fig:3a} (b). The EPG is measured in the standard way by fitting the $|0\rangle$ state as described in the caption to Fig.~\ref{fig:1}. To measure the leakage rate per gate (LPG) we fit the $|2\rangle$ state RB data to the same type of RB curve $A r^m + B$ and the LPG is given by the expression $p=(1-r)B/N_g$~\cite{wood:2017}. In general there is a correction to the RB fit of the $|0\rangle$ data due to leakage, however, when EPG$\gg$LPG this correction is small~\cite{wood:2017}. A typical calibration curve for the DRAG parameter of the DRAGZ pulses is shown in Fig.~\ref{fig:3a} (c). An efficient proxy for the LPG is the averaged $|2\rangle$ population for a long Clifford sequence. 

\begin{figure}
\includegraphics[width=0.4\textwidth]{\figfolder{3}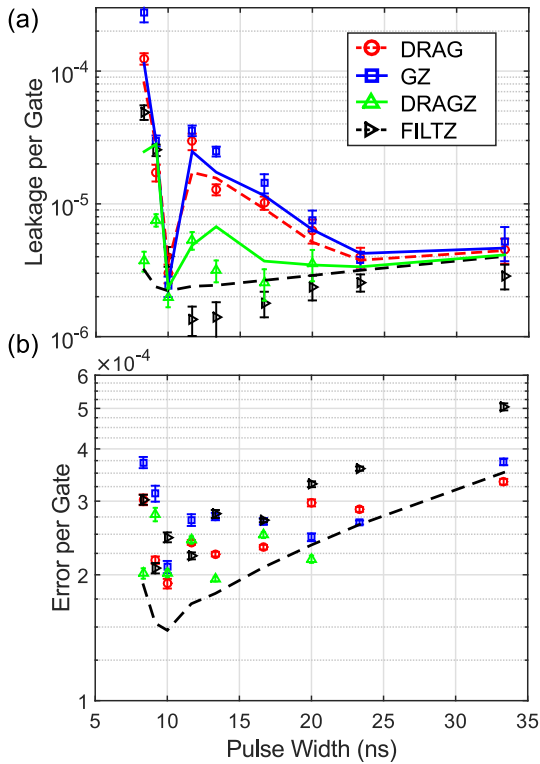}
\caption{(Color Online) (a) Leakage per gate versus pulse length for DRAG, GZ, DRAGZ and FILTZ. Lines are theory curves assuming the coherences mentioned in the main text, a temperature of 46~mK, LO leakage of -65~dBm and an AWG filter bandwidth of 300~MHz (100~MHz for FILTZ). DRAGZ points were only taken to a pulse length of 20~ns; beyond this point calibration of the DRAG parameter (to minimize leakage) was not reliable because the leakage signal was below the noise floor. (b) EPG from the same measurement. The black dashed line is a theory curve for a perfect DRAG pulse (i.e. no mixer or AWG effects) and is representative of the coherence limited error.  \label{fig:3}}
\end{figure}

The LPG and EPG versus pulse width for the four pulse types is illustrated in Fig.~\ref{fig:3}. DRAG and GZ give similar results with a general trend of lower LPG for longer pulse widths as expected due to Fourier broadening. The exception to this trend is a pronounced minima at 10~ns which is an artifact of the Gaussian truncation such that there is a zero in the pulse spectrum at $\omega_{12}$; this effect is captured accurately in the numerics. For pulses shorter than 20ns the DRAGZ and FILTZ pulses demonstrate nearly an order-of-magnitude lower LPG. For a pulse length of 13.3~ns we measure the LPG ($\times 10^{-6}$) for the various pulse types:  DRAGZ $3.1(6)$, FILTZ $1.4(4)$, DRAG $13(1)$ and GZ  $25(2)$. Overall, each of the pulses obtains similar EPG ($\times 10^{-4}$): DRAGZ $1.95(3)$, FILTZ $2.80(6)$, DRAG $2.24(3)$, GZ $2.75(6)$. These are all close to the coherence limit of $1.8 \times 10^{-4}$. While the LPG sets a lower bound on the EPG, significant gains in coherence will be required to reach that bound. For pulses greater than 20~ns the LPG rises; this is observed in theory calculations which include thermal relaxation to an effective system temperature of $T=46$~mK. When the system is at a finite temperature there is incoherent population transfer between levels $m$ and $n$ such that in equilibrium the ratio of populations is $e^{-(E_n-E_m)/k_B T}$. Therefore, finite-temperature heating sets a lower bound on the leakage, which is also the conclusion of Ref.~\cite{chen:2016}. Understanding why the qubit is higher temperature than the cryostat ($10-15$~mK) and how to reduce that temperature is an active area of investigation.

Overall, replacing DRAG with GZ pulses does not affect the EPG and instead it frees up DRAG pulse shaping to specifically minimize leakage using DRAGZ pulses. The lowest LPG is obtained with FILTZ, albeit with similar performance to DRAGZ and with the caveat that it only works for specific sideband frequencies. Ultimately, leakage is not a limiting factor for single qubit gate performance, however, leakage can have detrimental effects on error correction protocols~\cite{fowler:2013,suchara:2015}. 

\section{The VZ-Gate in Multiqubit Systems}

Employing VZ-gates in multiqubit systems depends on the specific implementation of the two-qubit gate interaction. For example, consider a two-qubit Hamiltonian with the interaction term $\hat{\sigma}^{(1)}_Z \otimes \hat{\sigma}^{(2)}_X$,
\begin{eqnarray}
	H/\hbar & = &  - \frac{\omega_{1}}{2} \hat{\sigma}^{(1)}_Z  - \frac{\omega_{2}}{2} \hat{\sigma}^{(2)}_Z +  g \hat{\sigma}^{(1)}_Z \otimes \hat{\sigma}^{(2)}_X. \label{eqn:int1}
\end{eqnarray}
In the rotating frame of the single-qubit drives, $U_{\mathrm{rot}} = e^{-i \left( \frac{\omega_{1}t + \phi_1}{2} \hat{\sigma}^{(1)}_Z  + \frac{\omega_{2}t + \phi_2}{2} \hat{\sigma}^{(2)}_Z \right)}$,
\begin{eqnarray}
\tilde{H}/\hbar & = &  g \hat{\sigma}^{(1)}_Z \otimes \left( \cos(\omega_2 t + \phi_2) \hat{\sigma}^{(2)}_X \right. \nonumber \\
					    & & \left. + \sin(\omega_2 t + \phi_2)\hat{\sigma}^{(2)}_Y \right), \label{eqn:int2}
\end{eqnarray} 
and so the single-qubit drive phase is imprinted on the two-qubit interaction if the interaction term is $\hat{\sigma}_X$ or $\hat{\sigma}_Y$. When we apply a VZ-gate, the phase update to the single-qubit drive will affect subsequent two-qubit interactions; how to manage this issue is dependent on the specific implementation of that interaction.  \\

For microwave-activated gates the phase update is straightforward. Here we will give two examples for the cross-resonance gate and the parametric iSWAP gate. The cross-resonance (CR) gate (see e.g. Ref~\cite{sheldon:2016}) interaction is the example considered in Eqn.~(\ref{eqn:int1}). To turn on the CR interaction the $ZX$ term is modulated at $\omega_2$ with a separate drive. The CR rotation angle is defined with respect to the the phase $\phi_2$ as shown in Eqn.~(\ref{eqn:int2}). When a VZ-gate is applied to qubit 2 the phase of the CR drive must be updated accordingly. For the iSWAP gate (see e.g. Ref.~\cite{mckay:2016}), an $XX+YY$ term is activated by modulating at the difference frequency $\omega_1-\omega_2$. The phase of this iSWAP drive is matched to the difference of the single-qubit drive phases $\phi_1-\phi_2$. Therefore, the VZ-gate phase is applied to the iSWAP drive with a different sign for qubits 1 and 2. 

For flux-tunable qubits, i.e., where the qubit frequencies are dynamically tuned to go to an interaction resonance, compatibility with the VZ-gate is more difficult. In these systems there are time dependent single-qubit $\sigma_Z$ terms in Eqn.~(\ref{eqn:int2}) which do not necessarily commute with the interaction. Therefore VZ-gates before the interaction necessitate also updating the $\sigma_Z$ dynamics. However, if the interaction is $ZZ$ (see e.g. Refs~\cite{dicarlo:2009,barends:2014}), then these single-qubit $Z$ terms commute through and can be compensated by a subsequent VZ-gate.  

\section{Conclusions}

In conclusion, we investigate a method to implement a near-perfect Z-gate by controlling the phase of the microwave drive used for $X$ and $Y$ rotations -- the virtual Z-gate (VZ-gate). This gate can improve the fidelity of circuits with a large number of single-qubit gates, can be used to efficiently correct typical gate errors and be used to implement arbitrary SU(2) gates given a calibrated $X_{\pi/2}$ gate. In this work we used VZ-gates to correct single-qubit rotation errors, but the gate should have wide applicability for improving two-qubit gates. In particular, as the number of qubits increases, crosstalk Z-errors will be ubiquitous. The VZ-gate is a low-overhead method for correcting these type of errors. By using pulse shaping techniques to minimize leakage and VZ gates to correct rotation errors we demonstrated a hybrid pulse with leakage limited by the qubit temperature and gate fidelity limited by coherence. Further improvements in leakage are limited by the qubit temperature.\\

\begin{acknowledgments}
We acknowledge Firat Solgun, George Keefe and Markus Brink for the simulation, layout and fabrication of devices. We acknowledge useful discussions with Lev Bishop. This work was supported by the Army Research Office under contract W911NF-14-1-0124.
\end{acknowledgments}

%

\end{document}